\documentclass[12pt,a4]{article}
\usepackage{amsmath,amssymb}
\usepackage{graphicx,psfrag,color}
\usepackage{epsfig}

\newcommand{\tr}{\operatorname{Tr}}

\newcommand{\ket}[1]{\left|#1\right\rangle}      
\newcommand{\bra}[1]{\left\langle #1\right|}     
\newcommand{\eq}{\begin{equation}}
\newcommand{\en}{\end{equation}}
\newcommand{\bear}{\begin{eqnarray}}
\newcommand{\ear}{\end{eqnarray}}
\newcommand{\bt} { \begin{tabular} }
\newcommand{\et}{ \end{tabular} }
\newcommand{\bc} { \begin{center} }
\newcommand{\ec}{ \end{center} }

\title{General scalar products in the arbitrary six-vertex model}

\author{G.A.P. Ribeiro\footnote{pavan@df.ufscar.br} \\  Departamento de F\'{i}sica, Universidade Federal de S\~ao Carlos \\ 13565-905 S\~ao Carlos-SP, Brazil}

\begin{document}
\maketitle

\begin{abstract}
In this work we use the algebraic Bethe ansatz to derive the general scalar product in the six-vertex model for generic Boltzmann weights. We performed this calculation using only the unitarity property, the Yang-Baxter algebra and the Yang-Baxter equation. We have derived a recurrence relation for the scalar product. The solution of this relation was written in terms of the domain wall partition functions. By its turn, these partition functions were also obtained for generic Boltzmann weights, which provided us with an explicit expression for the general scalar product.
\end{abstract}
\centerline{PACS: 75.10.Pq; 02.30.Ik;}
\centerline{Keywords: Algebraic Bethe ansatz, integrable systems, scalar products}

\thispagestyle{empty}

\newpage

\tableofcontents
\newpage

\section{Introduction}

It is well-known that the quantum inverse scattering method is a powerful tool to solve exactly quantum integrable models as well as their classical counterparts \cite{FADDEEV,KOREPIN-book}. Based on this approach one can construct the Bethe states by the action of pseudo-particle creation operators on a pseudo-vacuum state. These operators are the off-diagonal matrix elements of the monodromy matrix ${\cal T}_{\cal A}(\lambda)$. By its turn the monodromy matrix elements are the generators of the Yang-Baxter algebra given by the following quadratic relation,
\eq
R_{12}(\lambda, \mu) {\cal T}_{1}(\lambda) {\cal T}_{2}(\mu) = {\cal T}_{2}(\mu) {\cal T}_{1}(\lambda) R_{12}(\lambda, \mu).
\label{fundrel}
\en

The monodromy matrix elements act on the space of states of the quantum physical system and the $R$-matrix elements play the role of the structure constants of the above Yang-Baxter algebra (\ref{fundrel}). The latter algebra is associative thanks to the Yang-Baxter equation \cite{BAXTER}
\eq
R_{12}(\lambda,\mu)R_{13}(\lambda,\gamma)R_{23}(\mu,\gamma)=R_{23}(\mu,\gamma)R_{13}(\lambda,\gamma)R_{12}(\lambda,\mu).
\label{yangbaxter}
\en

The trace of the monodromy matrix over the auxiliary space $T(\lambda)=\tr_{\cal A}{\left[ {\cal T}_{\cal A}(\lambda) \right]}$ is the row-to-row transfer matrix. This matrix constitutes a family of commuting operators $\left[T(\lambda),T(\mu)\right]=0$, provided that the $R$-matrix is invertible. This condition is granted by the unitarity relation
\eq
R_{21}(\lambda,\mu)R_{12}(\mu,\lambda)= I,
\label{unitarity}
\en
where $I$ is the identity matrix.

Taking the action of the transfer matrix over the Bethe states, one is able to obtain the eigenvalues of the transfer matrix and its related conserved quantities. The eigenvalue expression is usually given in terms of some parameters which should satisfy the associated Bethe ansatz equations \cite{QISM-rev}. 

After obtaining the eigenvalues, one of the most challenging problems is to compute the scalar products and correlation functions. This was accomplished, by means of the algebraic Bethe ansatz, for models descending from the  $R$-matrix of the symmetric six-vertex model \cite{KOREPIN-norm,KOREPIN-IZERGIN,MAILLET}.

Recently, it was argued that the algebraic formulation of the Bethe states of the transfer matrix can be done using only the Yang-Baxter algebra (\ref{fundrel}), the Yang-Baxter equation (\ref{yangbaxter}) and the unitarity property (\ref{unitarity}) satisfied by the $R$-matrix \cite{MELO}. This way the on-shell properties (eigenvalues and the Bethe ansatz equations) as well as the off-shell  Bethe vectors are obtained in terms of arbitrary Boltzmann weights.

This idea have been shown to be valid also on the computation of the monodromy matrix elements of the arbitrary six-vertex model in the $F$-basis and the scalar product for arbitrary Boltzmann weights \cite{ZUPARIC}.

In this paper we shall obtain the scalar product as well as its associated domain wall partition functions by means of the algebraic Bethe ansatz for the completely asymmetric six-vertex model for arbitrary Boltzmann weights. The aim of this paper is to show that this can be accomplished using only the Yang-Baxter algebra, the Yang-Baxter equation and the unitarity relation.

The outline of the article is as follows. In section \ref{6ver}, we define the asymmetric six-vertex model and list the main relations needed in this work. In section \ref{SP}, we derive a recurrence relation for the scalar product and sketch its solution. In section \ref{DWPF}, we derive the partition function with domain wall boundary conditions which will help on writing a closed formula for the scalar product. Our conclusions are given on section \ref{CONCLUSION}.

\section{The six-vertex model}\label{6ver}

We define the $R$-matrix of the asymmetric six-vertex model as \cite{BAXTER,BAXTER2},
\eq
R(\lambda,\mu)=\left(\begin{array}{cccc}
	a_+(\lambda,\mu) & 0 & 0 &0 \\
	0 & b_+(\lambda,\mu) & c_+(\lambda,\mu) & 0 \\
	0 & c_-(\lambda,\mu) & b_-(\lambda,\mu) & 0 \\
	0 & 0 & 0 & a_-(\lambda,\mu) 
              \end{array}\right).
\label{r-matrix}
\en
This $R$-matrix should satisfy the unitarity property (\ref{unitarity}), which results on the following set of relations for its matrix elements
\bear
c_-(\lambda,\mu)b_-(\mu,\lambda)+b_-(\lambda,\mu)c_+(\mu,\lambda)&=&0,\label{unit1}\\
b_+(\lambda,\mu)c_-(\mu,\lambda)+c_+(\lambda,\mu)b_+(\mu,\lambda)&=&0,\label{unit2} \\
a_+(\lambda,\mu)a_+(\mu,\lambda)&=&1,  \label{unit3}\\
a_-(\lambda,\mu)a_-(\mu,\lambda)&=&1, \label{unit4} \\
c_-(\lambda,\mu)c_-(\mu,\lambda)+b_-(\lambda,\mu)b_+(\mu,\lambda)&=&1, \label{unit5}\\
c_+(\lambda,\mu)c_+(\mu,\lambda)+b_+(\lambda,\mu)b_-(\mu,\lambda)&=&1. \label{unit6}
\ear
Moreover the above $R$-matrix satisfies the Yang-Baxter equation (\ref{yangbaxter}). This provides us with an additional set of functional relations among the Boltzmann weights given by,
\bear
b_- c_-' b_+'' +c_- a_+' c_-'' &=& a_+ c_-' a_+'' \label{eq1},\\
c_- a_+' b_-'' +b_- c_-' c_+'' &=& c_- b_-' a_+''\label{eq2},\\
b_+ a_+' c_-'' +c_+ c_-' b_+'' &=& a_+ b_+' c_-''\label{eq3},
\ear
\bear
c_+ a_+' b_-'' +b_- c_+' c_-'' &=& c_+ b_-' a_+''\label{eq4},\\
b_+ c_-' b_-'' +c_- a_-' c_-'' &=& a_- c_-' a_-''\label{eq5},\\
b_- a_-' c_-'' +c_+ c_-' b_-'' &=& a_- b_-' c_-''\label{eq6},\\
b_+ a_+' c_+'' +c_- c_+' b_+'' &=& a_+ b_+' c_+''\label{eq7},\\
b_- c_+' b_+'' +c_+ a_+' c_+'' &=& a_+ c_+' a_+''\label{eq8},\\
c_- a_-' b_+'' +b_+ c_-' c_+'' &=& c_- b_+' a_-''\label{eq9},\\
b_- a_-' c_+'' +c_- c_+' b_-'' &=& a_- b_-' c_+''\label{eq10},\\
c_+ a_-' b_+'' +b_+ c_+' c_-'' &=& c_+ b_+' a_-''\label{eq11},\\
b_+ c_+' b_-'' +c_+ a_-' c_+'' &=& a_- c_+' a_-''\label{eq12},
\ear
where $a_{\pm}=a_{\pm}(\lambda,\mu)$, $a_{\pm}'=a_{\pm}(\lambda,\gamma)$ and $a_{\pm}''=a_{\pm}(\mu,\gamma)$ and likewise for the other functions.

The monodromy matrix ${\cal T}_{\cal A}(\lambda,\{\nu_k\})=R_{{\cal A}L}(\lambda,\nu_L)\cdots R_{{\cal A}1}(\lambda,\nu_1)$ satisfies the fundamental relation (\ref{fundrel}) thanks to the Yang-Baxter equation (\ref{yangbaxter}). For the six-vertex model, the monodromy matrix can be written as a $2\times 2$ matrix with operator valued entries,
\eq
{\cal T}(\lambda,\{\nu_k\})=\left(\begin{array}{cc}
	A(\lambda,\{\nu_k\}) & B(\lambda,\{\nu_k\}) \\
	C(\lambda,\{\nu_k\}) & D(\lambda,\{\nu_k\})
              \end{array}\right).
\label{monodromy}
\en
These operators have to obey sixteen commutation rules which follow from (\ref{fundrel}). We only list the commutation rules used in this work.
\bear
B(\lambda)B(\mu)&=&\frac{a_-(\lambda,\mu)}{a_+(\lambda,\mu)}B(\mu)B(\lambda), \label{alg1} \\
A(\lambda)B(\mu)&=&\frac{a_+(\mu,\lambda)}{b_-(\mu,\lambda)} B(\mu)A(\lambda) - \frac{c_+(\mu,\lambda)}{b_-(\mu,\lambda)} B(\lambda)A(\mu),\label{alg2} \\
D(\lambda)B(\mu)&=&\frac{a_-(\lambda,\mu)}{b_-(\lambda,\mu)} B(\mu)D(\lambda) - \frac{c_-(\lambda,\mu)}{b_-(\lambda,\mu)} B(\lambda)D(\mu),\label{alg3}
\ear
\bear
C(\lambda)C(\mu)&=&\frac{a_-(\mu,\lambda)}{a_+(\mu,\lambda)}C(\mu)C(\lambda),\label{alg4} \\
C(\lambda)A(\mu)&=&\frac{a_+(\lambda,\mu)}{b_-(\lambda,\mu)} A(\mu)C(\lambda) - \frac{c_-(\lambda,\mu)}{b_-(\lambda,\mu)} A(\lambda)C(\mu),\label{alg5} \\
C(\lambda)D(\mu)&=&\frac{a_-(\mu,\lambda)}{b_-(\mu,\lambda)} D(\mu)C(\lambda) - \frac{c_+(\mu,\lambda)}{b_-(\mu,\lambda)} D(\lambda)C(\mu),\label{alg6} \\
C(\lambda)B(\mu)&=&\frac{b_+(\lambda,\mu)}{b_-(\lambda,\mu)} B(\mu)C(\lambda) + \frac{c_-(\lambda,\mu)}{b_-(\lambda,\mu)} (A(\mu)D(\lambda)- A(\lambda)D(\mu) ).\label{alg7}
\ear

Additionally, the structure of the $R$-matrix (\ref{r-matrix}) implies that both ferromagnetic states $\ket{\Uparrow}=\bigotimes_{j=1}^L \left(\begin{array}{c} 1 \\ 0\end{array}\right)_j$ and $\ket{\Downarrow}=\bigotimes_{j=1}^L \left(\begin{array}{c} 0 \\ 1\end{array}\right)_j$ are innate eigenstates of the transfer matrix $T(\lambda,\{\nu_k\})=A(\lambda,\{\nu_k\})+D(\lambda,\{\nu_k\})$. Therefore, they play the role of pseudo-vacuum states.

One can see immediately that the action of the monodromy matrix over the pseudo-vacuum state, e.g.  $\ket{\Uparrow}$, results in a triangular matrix,
\eq
{\cal T}(\lambda,\{\nu_k\})\ket{\Uparrow}=\left(\begin{array}{cc}
	a(\lambda,\{\nu_k\})\ket{\Uparrow} & \# \\
	0 & d(\lambda,\{\nu_k\})\ket{\Uparrow}
              \end{array}\right),
\label{triang}
\en
where $a(\lambda,\{\nu_k\})$ and $d(\lambda,\{\nu_k\})$ correspond to some fixed representation and the symbol $\#$ stands for a non-null state. Therefore, one sees that the $B(\lambda,\{\nu_k\})$ and $C(\lambda,\{\nu_k\})$ operators are the creation and annihilation operators over the state $\ket{\Uparrow}$ \cite{QISM-rev}, which means
\begin{alignat}{2}
A(\lambda,\{\nu_k\})\ket{\Uparrow} &=a(\lambda,\{\nu_k\})\ket{\Uparrow} & \qquad B(\lambda,\{\nu_k\})\ket{\Uparrow} &=\#,   \\
C(\lambda,\{\nu_k\})\ket{\Uparrow} &=0  & \qquad D(\lambda,\{\nu_k\})\ket{\Uparrow}&=d(\lambda,\{\nu_k\})\ket{\Uparrow}, 
\end{alignat}
or alternatively,
\begin{alignat}{2}
\bra{\Uparrow} A(\lambda,\{\nu_k\}) &= \bra{\Uparrow} a(\lambda,\{\nu_k\}) & \qquad \bra{\Uparrow} B(\lambda,\{\nu_k\}) &=0,  \label{UPB} \\
\bra{\Uparrow} C(\lambda,\{\nu_k\}) &=\#  & \qquad \bra{\Uparrow} D(\lambda,\{\nu_k\}) &=\bra{\Uparrow} d(\lambda,\{\nu_k\}).
\end{alignat}

The action of the creation operators over the pseudo-vacuum is the algebraic version of the famous Bethe ansatz and is given by
\eq
\ket{\Psi_M}=B(\lambda_M,\{\nu_k\})\cdots B(\lambda_1,\{\nu_k\})\ket{\Uparrow}.
\label{KETstate}
\en

In order to obtain the eigenvalues of the transfer matrix one has to commute the operators $A(\lambda,\{\nu_k\})$ and $D(\lambda,\{\nu_k\})$ with $B(\lambda,\{\nu_k\})$. In doing so we need to use the relations  (\ref{eq10},\ref{alg1},\ref{alg2}) for the operator $A(\lambda,\{\nu_k\})$ and the relations (\ref{eq2},\ref{alg1},\ref{alg3}) for the operator $D(\lambda,\{\nu_k\})$\cite{MELO}, which results (see appendix A)
\begin{alignat}{1}
&A(\lambda,\{\nu_k\})\prod_{i=1}^M B(\lambda_i,\{\nu_k\})=\prod_{i=1}^M\frac{a_+(\lambda_i,\lambda)}{b_-(\lambda_i,\lambda)} \prod_{i=1}^M B(\lambda_i,\{\nu_k\}) A(\lambda,\{\nu_k\}) \nonumber\\
&- \sum_{j=1}^M\frac{c_+(\lambda_j,\lambda)}{b_-(\lambda_j,\lambda)} \prod_{\stackrel{i=1}{i\neq j}}^M \frac{a_+^{(\theta)}(\lambda_i,\lambda_j)}{b_-(\lambda_i,\lambda_j)} B(\lambda,\{\nu_k\}) \prod_{\stackrel{i=1}{i\neq j}}^M B(\lambda_i,\{\nu_k\}) A(\lambda_j,\{\nu_k\}), \label{AoverB} \\
&D(\lambda,\{\nu_k\})\prod_{i=1}^M B(\lambda_i,\{\nu_k\})=\prod_{i=1}^M\frac{a_-(\lambda,\lambda_i)}{b_-(\lambda,\lambda_i)} \prod_{i=1}^M B(\lambda_i,\{\nu_k\}) D(\lambda,\{\nu_k\}) \nonumber\\
&- \sum_{j=1}^M\frac{c_-(\lambda,\lambda_j)}{b_-(\lambda,\lambda_j)} \prod_{\stackrel{i=1}{i\neq j}}^M \frac{a_+^{(\theta)}(\lambda_j,\lambda_i)}{b_-(\lambda_j,\lambda_i)} B(\lambda,\{\nu_k\}) \prod_{\stackrel{i=1}{i\neq j}}^M B(\lambda_i,\{\nu_k\}) D(\lambda_j,\{\nu_k\}), \label{DoverB}
\end{alignat}
where $a_+^{(\theta)}(\lambda_i,\lambda_j)=a_+(\lambda_i,\lambda_j)\theta_{>}(\lambda_i,\lambda_j)$ and
\eq
\theta_>(\lambda_i,\lambda_j)=\begin{cases}
                               \frac{a_-(\lambda_i,\lambda_j)}{a_+(\lambda_i,\lambda_j)}, & i>j, \\
				1, & i\leq j.
                              \end{cases}
\en

Therefore, we obtain
\bear
T(\lambda)\ket{\Psi_M}=\Lambda_M(\lambda,\{\nu_k\}) \ket{\Psi_M} + \sum_{j=1}^M \Gamma_{j}(\lambda) B(\lambda,\{\nu_k\}) \prod_{\stackrel{i=1}{i\neq j}}^M B(\lambda_i,\{\nu_k\}) \ket{\Uparrow},
\label{eigen}
\ear
where the eigenvalues $\Lambda_M(\lambda,\{\nu_k\})$ are given by
\bear
\Lambda_M(\lambda,\{\nu_k\})=a(\lambda,\{\nu_k\})\prod_{i=1}^M\frac{a_+(\lambda_i,\lambda)}{b_-(\lambda_i,\lambda)} 
+d(\lambda,\{\nu_k\})\prod_{i=1}^M\frac{a_-(\lambda,\lambda_i)}{b_-(\lambda,\lambda_i)},
\ear
and
\bear
\Gamma_{j}(\lambda)&=&a(\lambda_j,\{\nu_k\}) \frac{c_+(\lambda_j,\lambda)}{b_-(\lambda_j,\lambda)} \prod_{\stackrel{i=1}{i\neq j}}^M \frac{a_+^{(\theta)}(\lambda_i,\lambda_j)}{b_-(\lambda_i,\lambda_j)}  \\
&+&d(\lambda_j,\{\nu_k\})\frac{c_-(\lambda,\lambda_j)}{b_-(\lambda,\lambda_j)} \prod_{\stackrel{i=1}{i\neq j}}^M \frac{a_+^{(\theta)}(\lambda_j,\lambda_i)}{b_-(\lambda_j,\lambda_i)}. \nonumber
\ear

Finally, we have to impose the coefficient $\Gamma_j(\lambda)$ to vanish for arbitrary $\lambda$ in order to cancel the unwanted terms. Using the unitarity relation (\ref{unit1}), one sees that the parameters $\lambda_j$ have to satisfy the Bethe ansatz equations
\eq
\frac{a(\lambda_j,\{\nu_k\})}{d(\lambda_j,\{\nu_k\})}=\prod_{\stackrel{i=1}{i\neq j}}^M \frac{a_+^{(\theta)}(\lambda_j,\lambda_i)}{a_+^{(\theta)}(\lambda_i,\lambda_j)}\frac{b_-(\lambda_i,\lambda_j)}{b_-(\lambda_j,\lambda_i)}.
\label{BAeq}
\en

Similarly the dual Bethe state can be written as follow,
\eq
\bra{\Psi_M}=\bra{\Uparrow}C(\lambda_1,\{\nu_k\})\cdots C(\lambda_M,\{\nu_k\}).
\label{BRAstate}
\en
However, in order to obtain the transfer matrix eigenvalues one should look for commutation rules between the operators $A(\lambda,\{\nu_k\})$ and $D(\lambda,\{\nu_k\})$ with $C(\lambda,\{\nu_k\})$. Here, one has to use the relations (\ref{eq6},\ref{alg4},\ref{alg5}) for the operator $A(\lambda,\{\nu_k\})$ and the relations (\ref{eq4},\ref{alg4},\ref{alg6}) for the operator $D(\lambda,\{\nu_k\})$, such that
\begin{alignat}{1}
&\prod_{i=1}^M C(\lambda_i,\{\nu_k\}) A(\lambda,\{\nu_k\})=\prod_{i=1}^M\frac{a_+(\lambda_i,\lambda)}{b_-(\lambda_i,\lambda)} A(\lambda,\{\nu_k\}) \prod_{i=1}^M  C(\lambda_i,\{\nu_k\})  \nonumber\\
&- \sum_{j=1}^M\frac{c_-(\lambda_j,\lambda)}{b_-(\lambda_j,\lambda)} \prod_{\stackrel{i=1}{i\neq j}}^M \frac{a_+^{(\theta)}(\lambda_i,\lambda_j)}{b_-(\lambda_i,\lambda_j)} A(\lambda_j,\{\nu_k\}) \prod_{\stackrel{i=1}{i\neq j}}^M C(\lambda_i,\{\nu_k\}) C(\lambda,\{\nu_k\}), \label{CMA} \\
&\prod_{i=1}^M C(\lambda_i,\{\nu_k\})D(\lambda,\{\nu_k\})=\prod_{i=1}^M\frac{a_-(\lambda,\lambda_i)}{b_-(\lambda,\lambda_i)} D(\lambda,\{\nu_k\})\prod_{i=1}^M C(\lambda_i,\{\nu_k\})  \nonumber\\
&- \sum_{j=1}^M\frac{c_+(\lambda,\lambda_j)}{b_-(\lambda,\lambda_j)} \prod_{\stackrel{i=1}{i\neq j}}^M \frac{a_+^{(\theta)}(\lambda_j,\lambda_i)}{b_-(\lambda_j,\lambda_i)}  D(\lambda_j,\{\nu_k\})\prod_{\stackrel{i=1}{i\neq j}}^M C(\lambda_i,\{\nu_k\}) C(\lambda,\{\nu_k\}). \label{CMD}
\end{alignat}

\section{The scalar product}\label{SP}

The general scalar product is defined as
\eq
S_M(\{\lambda\}_{i=1}^M|\{\mu\}_{i=1}^M)=\bra{\Uparrow} C(\mu_1)\cdots C(\mu_M) B(\lambda_M)\cdots B(\lambda_1) \ket{\Uparrow} 
\label{genSP}
\en
where we have dropped the dependence on $\nu_k$. In the particular case that both set of parameters $\lambda_j $ and $\mu_j$ satisfy the Bethe ansatz equations (\ref{BAeq}), the product (\ref{genSP}) becomes the norm of Bethe eigenstates\cite{KOREPIN-norm}.

To compute the scalar product (\ref{genSP}), we must calculate the action of the operator $B(\lambda)$ over the state (\ref{BRAstate}) (or alternatively, the action of $C(\mu)$ over the state (\ref{KETstate})). To this purpose we have to use repeatedly the relation (\ref{alg7}), which results
\bear
\bra{\Uparrow} C(\mu_1)\cdots C(\mu_M) B(\lambda_M)=\prod_{i=1}^M \frac{b_+(\mu_i,\lambda_M)}{b_-(\mu_i,\lambda_M)} \bra{\Uparrow} B(\lambda_M)C(\mu_1)\cdots C(\mu_M) \nonumber\\
+ \sum_{j=1}^M \frac{c_-(\mu_j,\lambda_M)}{b_-(\mu_j,\lambda_M)} \bra{\Uparrow}\prod_{i=1}^{j-1}C(\mu_i) \left( A(\lambda_M) D(\mu_j) -A(\mu_j)D(\lambda_M) \right) \prod_{i=j+1}^M C(\mu_i),
\label{CMB}
\ear
where the first term vanishes thanks to (\ref{UPB}) and we set $\lambda_i=\mu_{2M +1-i}$, $i=1,\cdots,M$ for convenience.

After that, one has to compute the action of $A(\lambda)D(\mu_j)-A(\mu_j)D(\lambda)$ over the $C(\mu_i)$, where one should use the algebraic relations (\ref{CMA}-\ref{CMD}) along the same lines as \cite{KOREPIN-book,KOREPIN-norm}, such that
\begin{alignat}{1}
&\bra{\Uparrow} C(\mu_1)\cdots C(\mu_M) B(\mu_{M+1})= \sum_{\stackrel{j,k=1}{j\neq k}}^{M+1} \bra{\Uparrow}A(\mu_k)D(\mu_j) \times \\
&\frac{c_-(\mu_j,\mu_{M+1})c_+(\mu_{M+1},\mu_k)a_+^{(\theta)}(\mu_j,\mu_k)}{a_+^{(\theta)}(\mu_j,\mu_{M+1}) a_+^{(\theta)}(\mu_{M+1},\mu_k) b_-(\mu_j,\mu_k) } \prod_{\stackrel{i=1}{\stackrel{i\neq j}{i\neq k}}}^{M+1} \frac{a_+^{(\theta)}(\mu_i,\mu_k)}{b_-(\mu_i,\mu_k)}\frac{a_+^{(\theta)}(\mu_j,\mu_i)}{b_-(\mu_j,\mu_i)} \prod_{\stackrel{i=1}{\stackrel{i\neq j}{i\neq k}}}^{M+1}C(\mu_i), \nonumber
\end{alignat}
where we have also used the functional relation (\ref{eq6}) and the unitarity relations (\ref{unit3}-\ref{unit5}) for simplifications.

So we multiply the previous expression by additional $B(\mu_i)$ operators, $S_M(\{\mu\}_{i=M+1}^{2M}|\{\mu\}_{i=1}^M)=\bra{\Uparrow} C(\mu_1)\cdots C(\mu_M) B(\mu_{M+1})\prod_{i=M+2}^{2M} B(\mu_i) \ket{\Uparrow}$. This results in a recurrence relation for the scalar product given by,
\begin{alignat}{1}
&S_M(\{\mu\}_{i=M+1}^{2M}|\{\mu\}_{i=1}^M)=\sum_{\stackrel{j,k=1}{j\neq k}}^{M+1} \frac{a(\mu_k)d(\mu_j)c_-(\mu_j,\mu_{M+1})c_+(\mu_{M+1},\mu_k)a_+^{(\theta)}(\mu_j,\mu_k)}{a_+^{(\theta)}(\mu_j,\mu_{M+1}) a_+^{(\theta)}(\mu_{M+1},\mu_k) b_-(\mu_j,\mu_k) } \times \nonumber\\
&\times \prod_{\stackrel{i=1}{\stackrel{i\neq j}{i\neq k}}}^{M+1} \frac{a_+^{(\theta)}(\mu_i,\mu_k)}{b_-(\mu_i,\mu_k)}\frac{a_+^{(\theta)}(\mu_j,\mu_i)}{b_-(\mu_j,\mu_i)} S_{M-1}(\{\mu\}_{i=M+2}^{2M}|\{\mu\}_{\stackrel{i=1}{\stackrel{i\neq j}{i\neq k}}}^M).
\label{recurrence}
\end{alignat}

One can iterate the above recursion relation (\ref{recurrence}) and represent the resulting expression as follows,
\bear
S_M(\{\lambda\}|\{\mu\})=\sum_{\{\lambda\}=\{\lambda^+\}\cup \{\lambda^-\} \atop {\{\mu\}=\{\mu^+\}\cup \{\mu^-\} \atop |\lambda^+ |+ |\mu^-|=n_+} } \prod_{i=1}^{n_+} a(\lambda_i^+)d(\mu_i^+) \prod_{i=1}^{n_-} a(\mu_i^-)d(\lambda_i^-) K_M\left({\{\lambda^+\} \atop \{\lambda^-\}}\Big|{\{\mu^+\} \atop \{\mu^-\}}\right),
\label{combinat}
\ear
where we sum with respect to all partitions of the set $\{\lambda\}\cup\{\mu\}$  into two disjoint sets $\{\lambda^+\}\cup\{\mu^-\}$
 and $\{\lambda^-\}\cup\{\mu^+\}$ of $M$ of elements. If the number of elements in the sets $\{\lambda^{+}\}$ and $\{\mu^{+}\}$ is $n_{+}=|\lambda^{+}|=|\mu^{+}|$, then we have $n_-=|\lambda^{-}|=|\mu_-|=M-n_+$ elements in the sets $\{\lambda^{-}\}$ and $\{\mu^{-}\}$. 

Notice that any coefficient $K_M$ is determined only by terms arising from algebraic relations among monodromy matrix elements. Therefore, $K_M$ is independent of the representation, i.e. independent of the choice of the functions $a(\lambda,\{\nu_k\})$ and $d(\lambda,\{\nu_k\})$ \cite{KOREPIN-book}.

In view of that one can fix the coefficients $K_M$ using any special representation. Let's consider the case where $L=M$ and 
\begin{alignat}{2}
&a_M(\lambda)=\prod_{i=1}^M a_+(\lambda,\nu_i), & \qquad d_M(\lambda)=\prod_{i=1}^M b_-(\lambda,\nu_i),
\end{alignat}
where we have fixed the inhomogeneities as follows,
\bear
\nu_i=\begin{cases}
       \lambda_i^+, & i=1,\cdots,n_+ \\
	\mu_{i-n_+}^-, & i=n_+ +1,\cdots,M.
      \end{cases}
\label{inhomoge}
\ear 

Using the fact that the unitarity relation (\ref{unit1}) implies that $b_-(\lambda,\lambda)=0$, one sees that $d_M(\lambda)=0$ for any $\lambda \in \{\lambda^+\}\cup\{\mu^-\}$ and $d_M(\lambda)\neq 0$ for any $\lambda \in \{\lambda^-\}\cup\{\mu^+\}$. Consequently, there is only one non-vanishing term in the sum (\ref{combinat}), which allow us to write
\bear
\prod_{i=1}^{n_+} a_M(\lambda_i^+)d_M(\mu_i^+) \prod_{i=1}^{n_-} a_M(\mu_i^-)d_M(\lambda_i^-) K_M(\{\lambda^+\},\{\lambda^-\}|\{\mu^+\},\{\mu^-\}) \nonumber\\
=  \bra{\Uparrow_M} C(\mu_1)\cdots C(\mu_M) B(\lambda_M)\cdots B(\lambda_1) \ket{\Uparrow_M},
\ear
where $\ket{\Uparrow_M}$ is a state with $M$ spins up.

The product of $B$-operators overturns all $M$ spins resulting in the state $\ket{\Downarrow_M}$. Therefore, we have
\bear
\bra{\Uparrow_M} C(\mu_1)\cdots C(\mu_M) B(\lambda_M)\cdots B(\lambda_1) \ket{\Uparrow_M}= \nonumber\\
=  Z_M^{(C)}(\{\mu\};\{\lambda^+\}\cup \{\mu_-\}) Z_M^{(B)}(\{\lambda\};\{\lambda^+\}\cup \{\mu_-\}),
\ear
where the functions $Z_M^{(B,C)}$ are defined by
\bear 
Z_M^{(B)}(\{\lambda\};\{\lambda^+\}\cup \{\mu_-\})&=&\bra{\Downarrow_M} B(\lambda_M)\cdots B(\lambda_1) \ket{\Uparrow_M}, \label{ZB} \\
Z_M^{(C)}(\{\mu\};\{\lambda^+\}\cup \{\mu_-\})&=&\bra{\Uparrow_M} C(\mu_1)\cdots C(\mu_M)\ket{\Downarrow_M}. \label{ZC}
\ear
Finally, the coefficient $K_M$ can be written in terms of the above functions as follows
\bear
K_M\left({\{\lambda^+\} \atop \{\lambda^-\}}\Big|{\{\mu^+\} \atop \{\mu^-\}}\right)=\frac{Z_M^{(C)}(\{\mu\};\{\lambda^+\}\cup \{\mu_-\}) Z_M^{(B)}(\{\lambda\};\{\lambda^+\}\cup \{\mu_-\})}{\prod_{i=1}^{n_+} a_M(\lambda_i^+)d_M(\mu_i^+) \prod_{i=1}^{n_-} a_M(\mu_i^-)d_M(\lambda_i^-)}.
\label{KN}
\ear

The functions (\ref{ZB}-\ref{ZC}) are usually called domain wall partition function \cite{KOREPIN-norm} and play an important role on the calculation of scalar products and correlation function. We shall compute these partition functions on the next section in order to write a closed formula for the scalar product (\ref{combinat}).

\section{The domain wall partition function}\label{DWPF}

In this section we derive a recurrence relation for the partition function for the asymmetric six-vertex model with domain wall boundary conditions for arbitrary Boltzmann weights. This way we will proceed along the same lines of \cite{PRONKO} and define a couple of auxiliary one-point boundary correlation functions as follows,
\bear
G_{M,N}^{(B)}=\frac{1}{Z_{M}^{(B)}}\bra{\Downarrow_M} B(\lambda_M)\cdots B(\lambda_{N+1})p_1^{-}B(\lambda_N)B(\lambda_{N-1}) \cdots B(\lambda_1) \ket{\Uparrow_M}, \\
H_{M,N}^{(B)}=\frac{1}{Z_{M}^{(B)}}\bra{\Downarrow_M} B(\lambda_M)\cdots B(\lambda_{N+1})p_1^{-}B(\lambda_N)p_1^{+}B(\lambda_{N-1})\cdots B(\lambda_1) \ket{\Uparrow_M}, 
\ear
where $p_{1}^{\pm}=\frac{1}{2}(1\pm \sigma_1^z)$ are the local spin up and down projectors and $G_{M,M}^{(B)}=1$. The above boundary correlations are related by
\bear
G_{M,N}^{(B)}&=&\sum_{j=1}^N H_{M,j}^{(B)}, \\
G_{M,N}^{(B)}&=&H_{M,N}^{(B)} + G_{M,N-1}^{(B)}.
\ear

One can use the above relations in order to write the domain wall partition function $Z_{M}^{(B)}$ in terms of the boundary correlations
\bear
Z_{M}^{(B)}(\{\lambda\};\{\nu\})=\sum_{j=1}^{M}\bra{\Downarrow_M} B(\lambda_M)\cdots B(\lambda_{j+1})p_1^{-}B(\lambda_j)p_1^{+}B(\lambda_{j-1})\cdots B(\lambda_1) \ket{\Uparrow_M}.
\label{ZSUMH}
\ear

To derive the recurrence relation for the partition function we use the two-site model decomposition. More specifically, one has to decompose the monodromy matrix into two parts and introduce two monodromy matrices
\bear
{\cal T}_{\cal A}(\lambda)={\cal T}_{\cal A}^{(2)}(\lambda){\cal T}_{\cal A}^{(1)}(\lambda)=\left(\begin{array}{cc}
                                                       A(\lambda) & B(\lambda) \\
						       C(\lambda) & D(\lambda)
                                                      \end{array}\right),
\ear
such that
\bear
{\cal T}_{\cal A}^{(1)}=R_{{\cal A}1}(\lambda,\nu_1) =\left(\begin{array}{cc}
                                                       A_1(\lambda) & B_1(\lambda) \\
						       C_1(\lambda) & D_1(\lambda)
                                                      \end{array}\right), \\
{\cal T}_{\cal A}^{(2)}=R_{{\cal A}M}(\lambda,\nu_M)\cdots R_{{\cal A}2}(\lambda,\nu_2)=\left(\begin{array}{cc}
                                                       A_2(\lambda) & B_2(\lambda) \\
						       C_2(\lambda) & D_2(\lambda)
                                                      \end{array}\right),
\ear
which act on the states $\ket{\Uparrow_i}$, $i=1,2$ given by
\bear
\ket{\Uparrow_1}=\left(\begin{array}{c} 1 \\ 0\end{array}\right)_1, & \qquad \ket{\Uparrow_2}=\bigotimes_{j=2}^M \left(\begin{array}{c} 1 \\ 0\end{array}\right)_j .
\ear

In particular, the required monodromy matrix elements are readily obtained  
\bear 
B(\lambda)&=&A_2(\lambda)B_1(\lambda) + B_2(\lambda) D_1(\lambda), \label{Bglobal} \\
B_1(\lambda)&=&c_+(\lambda,\nu_1) \sigma_1^-, \\
D_1(\lambda)&=&b_-(\lambda,\nu_1) p_1^+ + a_-(\lambda,\nu_1) p_1^- , \label{Dlocal1}
\ear
and the action of the operator $A_2(\lambda)$ on the state $\ket{\Uparrow_2}$  is given by
\eq
A_2(\lambda)\ket{\Uparrow_2}=\prod_{i=2}^M a_+(\lambda,\nu_i) \ket{\Uparrow_2}.
\en

Using (\ref{Bglobal}-\ref{Dlocal1}), we can rewrite (\ref{ZSUMH}) in terms of the operators $A_2(\lambda)$ and $B_2(\lambda)$ as follows,
\bear
Z_{M}^{(B)}(\{\lambda\};\{\nu\})&=&\sum_{j=1}^{M}c_+(\lambda_j,\nu_1) \prod_{i=1}^{j-1} b_-(\lambda_i,\nu_1) \prod_{i=j+1}^{M} a_-(\lambda_i,\nu_1)   \\
&\times&\bra{\Downarrow_2} B_2(\lambda_M)\cdots B_2(\lambda_{j+1})A_2(\lambda_j)B_2(\lambda_{j-1})\cdots B_2(\lambda_1) \ket{\Uparrow_2}. \nonumber
\ear
At this point one should use the algebraic relations (\ref{alg1},\ref{AoverB}) \cite{PRONKO} in order to obtain the recurrence relation for the domain wall partition function,
\bear
Z_{M}^{(B)}(\{\lambda\}_{i=1}^M;\{\nu\}_{i=1}^M)&=&\sum_{j=1}^{M}c_+(\lambda_j,\nu_1) \prod_{\stackrel{i=1}{i\neq j}}^{M} b_-(\lambda_i,\nu_1) \prod_{\stackrel{i=1}{i\neq 1}}^{M} a_+(\lambda_j,\nu_i) \nonumber\\
&\times & \prod_{\stackrel{i=1}{i\neq j}}^M \frac{a_+^{(\theta)}(\lambda_i,\lambda_j)}{b_-(\lambda_i,\lambda_j)} Z_{M-1}^{(B)}(\{\lambda\}_{\stackrel{i=1}{i\neq j}}^M;\{\nu\}_{\stackrel{i=1}{i\neq 1}}^M).
\label{ZBREC}
\ear

Iterating the above relation $M-1$ times, one gets the solution as a sum over all permutations $P$ of $\{\lambda\}_{i=1}^M$

\bear
Z_{M}^{(B)}(\{\lambda\}_{i=1}^M;\{\nu\}_{i=1}^M)=\sum_{P} \prod_{i=1}^M c_+(\lambda_{P_i},\nu_i) \prod_{\stackrel{i,j=1}{i<j}}^{M} \frac{a_+^{(\theta)}(\lambda_{P_i},\lambda_{P_j})}{b_-(\lambda_{P_i},\lambda_{P_j})} a_+(\lambda_{P_i},\nu_j) b_-(\lambda_{P_j},\nu_i). 
\label{ZBM}
\ear

We still have to determine the partition function $Z_M^{(C)}$. The calculation goes in a similar way as above. Therefore, we just present the final result for the recurrence relation,
\bear
Z_{M}^{(C)}(\{\mu\}_{i=1}^M;\{\nu\}_{i=1}^M)&=&\sum_{j=1}^{M}c_-(\mu_j,\nu_M) \prod_{\stackrel{i=1}{i\neq j}}^{M} b_-(\mu_i,\nu_M) \prod_{\stackrel{i=1}{i\neq M}}^{M} a_+(\mu_j,\nu_i) \nonumber\\
&\times & \prod_{\stackrel{i=1}{i\neq j}}^M \frac{a_+^{(\theta)}(\mu_i,\mu_j)}{b_-(\mu_i,\mu_j)} Z_{M-1}^{(C)}(\{\mu\}_{\stackrel{i=1}{i\neq j}}^M;\{\nu\}_{\stackrel{i=1}{i\neq M}}^M),
\label{ZCREC}
\ear
whose solution obtained by iteration is given as a sum over all permutations $\bar{P}$ of $\{\mu\}_{i=1}^M$
\bear
Z_{M}^{(C)}(\{\mu\}_{i=1}^M;\{\nu\}_{i=1}^M)=\sum_{\bar{P}} \prod_{i=1}^M c_-(\mu_{\bar{P}_i},\nu_i) \prod_{\stackrel{i,j=1}{i<j}}^{M} \frac{a_+^{(\theta)}(\mu_{\bar{P}_i},\mu_{\bar{P}_j})}{b_-(\mu_{\bar{P}_i},\mu_{\bar{P}_j})} a_+(\mu_{\bar{P}_i},\nu_j) b_-(\mu_{\bar{P}_j},\nu_i). 
\label{ZCM}
\ear

\section{Explicit expression}

Using the special choice for the inhomogeneities (\ref{inhomoge}) and the recurrence relations (\ref{ZBREC},\ref{ZCREC}), we can write a simplified expression for the coefficient $K_M$ as follows,
\bear
K_M\left({\{\lambda^+\} \atop \{\lambda^-\}}\Big|{\{\mu^+\} \atop \{\mu^-\}}\right)&=&\prod_{j=1}^{n_+}\prod_{k=1}^{n_-}\frac{a_+(\mu_j^+,\mu_k^-)}{b_-(\mu_j^+,\mu_k^-)} \frac{a_-(\lambda_k^-,\lambda_j^+)}{b_-(\lambda_k^-,\lambda_j^+)}\prod_{\stackrel{j,k=1}{j<k}}^{n_+}\frac{a_-(\lambda_j^+,\lambda_k^+)}{a_+(\lambda_j^+,\lambda_k^+)} \nonumber \\
&\times&  \frac{Z_{n_+}^{(C)}(\{\mu^+\};\{\lambda^+\}) Z_{n_-}^{(B)}(\{\lambda^-\};\{\mu^-\}) }{\prod_{j,k=1}^{n_+}b_-(\mu_j^+,\lambda_k^+) \prod_{j,k=1}^{n_-}b_-(\lambda_j^-,\mu_k^-)}.
\label{KMsimp}
\ear

Finally, if we substitute the expression (\ref{ZBM},\ref{ZCM},\ref{KMsimp}) in (\ref{combinat}), we obtain an explicit expression for the scalar product for arbitrary Boltzmann weights,
\bear
S_M(\{\lambda\}|\{\mu\})&=&\sum_{\{\lambda\}=\{\lambda^+\}\cup \{\lambda^-\} \atop {\{\mu\}=\{\mu^+\}\cup \{\mu^-\} \atop |\lambda^+ |+ |\mu^-|=n_+} } \frac{\prod_{i=1}^{n_+} a(\lambda_i^+)d(\mu_i^+) \prod_{i=1}^{n_-} a(\mu_i^-)d(\lambda_i^-)} {\prod_{j,k=1}^{n_+}b_-(\mu_j^+,\lambda_k^+)\prod_{j,k=1}^{n_-}b_-(\lambda_j^-,\mu_k^-)} \nonumber \\
&\times&\prod_{j=1}^{n_+}\prod_{k=1}^{n_-}\frac{a_+(\mu_j^+,\mu_k^-)}{b_-(\mu_j^+,\mu_k^-)} \frac{a_-(\lambda_k^-,\lambda_j^+)}{b_-(\lambda_k^-,\lambda_j^+)}\prod_{\stackrel{j,k=1}{j<k}}^{n_+}\frac{a_-(\lambda_j^+,\lambda_k^+)}{a_+(\lambda_j^+,\lambda_k^+)} \\
&\times&  
\sum_{\bar{P}} \prod_{i=1}^{n^+} c_-(\mu_{\bar{P}_i}^+,\lambda_i^+) \prod_{\stackrel{i,j=1}{i<j}}^{n_+} \frac{ a_+^{(\theta)}(\mu^+_{\bar{P}_i},\mu^+_{\bar{P}_j})}{b_-(\mu^+_{\bar{P}_i},\mu^+_{\bar{P}_j})} a_+(\mu^+_{\bar{P}_i},\lambda^+_j) b_-(\mu^+_{\bar{P}_j},\lambda^+_i) \nonumber\\
&\times& \sum_{P} \prod_{i=1}^{n_-} c_+(\lambda^-_{P_i},\mu^-_i) \prod_{\stackrel{i,j=1}{i<j}}^{n_-} \frac{a_+^{(\theta)}(\lambda^-_{P_i},\lambda^-_{P_j})}{b_-(\lambda^-_{P_i},\lambda^-_{P_j})} a_+(\lambda^-_{P_i},\mu^-_j) b_-(\lambda^-_{P_j},\mu^-_i). \nonumber
\ear

\section{Conclusion}
\label{CONCLUSION}

In this paper we obtained an explicit expression for the scalar product and the domain wall partition function for the arbitrary six-vertex model. 

We have reviewed the algebraic Bethe for the six-vertex model under the new perspective of working with arbitrary Boltzmann weights. Using the main ingredients of the algebraic Bethe ansatz, we have been able to derive a recurrence relation for the scalar product. We managed to write the solution of this recurrence relation in terms of the domain wall partition functions. Then, we obtained a closed formula for the required partition functions. Finally, we have obtained an explicit expression for the general scalar product in terms of the arbitrary Boltzmann weights. We have done all that only using the Yang-Baxter algebra, the unitarity relation and the Yang-Baxter relations.

\section*{Acknowledgments}
The author thanks F. G\"ohmann for introducing him to this topic and M.J. Martins for discussions. This work has been supported by FAPESP and CNPq.

\addcontentsline{toc}{section}{Appendix A}
\section*{\bf Appendix A: two and three-particle state}\label{appendix}
\setcounter{equation}{0}
\renewcommand{\theequation}{A.\arabic{equation}}
\setcounter{section}{0}
\renewcommand{\thesection}{A.\arabic{section}}

In this appendix we work out in detail the commutation of the operator $A(\lambda)$ with two and three $B$-operators in order to exemplify the use of Yang-Baxter relations (\ref{eq1}-\ref{eq12}). In fact, to pass the $A$-operator over any number of $B$'s one need only to use the Yang-Baxter relation (\ref{eq10}). 

\section{$M=2$}
Let's start with the two $B$-operators case and use the commutation rule (\ref{alg2}) twice, which results
\bear
&&A(\lambda)B(\lambda_2)B(\lambda_1)=\frac{a_+(\lambda_2,\lambda)}{b_-(\lambda_2,\lambda)}\frac{a_+(\lambda_1,\lambda)}{b_-(\lambda_1,\lambda)} B(\lambda_2)B(\lambda_1)A(\lambda)  \nonumber\\
&-& \frac{c_+(\lambda_2,\lambda)}{b_-(\lambda_2,\lambda)}\frac{a_+(\lambda_1,\lambda_2)}{b_-(\lambda_1,\lambda_2)}B(\lambda)B(\lambda_1)A(\lambda_2) \\
&-&\Big[\frac{a_+(\lambda_2,\lambda)}{b_-(\lambda_2,\lambda)}\frac{c_+(\lambda_1,\lambda)}{b_-(\lambda_1,\lambda)} \underbrace{B(\lambda_2)B(\lambda)}_{eq. (\ref{alg1})} - \frac{c_+(\lambda_2,\lambda)}{b_-(\lambda_2,\lambda)}\underbrace{\frac{c_+(\lambda_1,\lambda_2)}{b_-(\lambda_1,\lambda_2)}}_{eq. (\ref{unit1})} B(\lambda)B(\lambda_2) \Big]A(\lambda_1).  \nonumber
\ear
We can further manipulate the third term of the above expression using the commutation rule (\ref{alg1}) and the unitarity relation (\ref{unit1}) as indicated. This provides us with the following expression	
\begin{alignat}{1}
&A(\lambda)B(\lambda_2)B(\lambda_1)=\frac{a_+(\lambda_2,\lambda)}{b_-(\lambda_2,\lambda)}\frac{a_+(\lambda_1,\lambda)}{b_-(\lambda_1,\lambda)}  B(\lambda_2)B(\lambda_1) A(\lambda) \nonumber \\
&- \frac{c_+(\lambda_2,\lambda)}{b_-(\lambda_2,\lambda)}\frac{a_+(\lambda_1,\lambda_2)}{b_-(\lambda_1,\lambda_2)}B(\lambda)B(\lambda_1)A(\lambda_2) \label{A2} \\
&-\Big[\frac{\overbrace{b_-(\lambda_2,\lambda_1) a_-(\lambda_2,\lambda) c_+(\lambda_1,\lambda) + c_-(\lambda_2,\lambda_1)  c_+(\lambda_2,\lambda) b_-(\lambda_1,\lambda)}^{eq. (\ref{eq10})}}{b_-(\lambda_2,\lambda_1)b_-(\lambda_2,\lambda)b_-(\lambda_1,\lambda)}  \Big] B(\lambda)B(\lambda_2)A(\lambda_1).  \nonumber
\end{alignat}
Finally, one can see that the resulting expression for the third term coincides with the left hand side of the Yang-Baxter relation (\ref{eq10}). After substituting (\ref{eq10}) in (\ref{A2}), we obtain
\begin{alignat}{1}
A(\lambda)B(\lambda_2)B(\lambda_1)&=\frac{a_+(\lambda_2,\lambda)}{b_-(\lambda_2,\lambda)}\frac{a_+(\lambda_1,\lambda)}{b_-(\lambda_1,\lambda)}  B(\lambda_2)B(\lambda_1) A(\lambda) \nonumber \\
&- \frac{c_+(\lambda_2,\lambda)}{b_-(\lambda_2,\lambda)}\frac{a_+(\lambda_1,\lambda_2)}{b_-(\lambda_1,\lambda_2)}B(\lambda)B(\lambda_1)A(\lambda_2) \label{A3} \\
&-\frac{c_+(\lambda_1,\lambda) }{b_-(\lambda_1,\lambda)}\frac{a_-(\lambda_2,\lambda_1)}{b_-(\lambda_2,\lambda_1)} B(\lambda)B(\lambda_2)A(\lambda_1).  \nonumber
\end{alignat}

\section{$M=3$}
Now we turn to the case where we have three $B$-operators. We use the previous result (\ref{A3}) and the commutation rule (\ref{alg1}), such that
\begin{alignat}{1}
&A(\lambda)B(\lambda_3)B(\lambda_2)B(\lambda_1)=\frac{a_+(\lambda_3,\lambda)}{b_-(\lambda_3,\lambda)}\frac{a_+(\lambda_2,\lambda)}{b_-(\lambda_2,\lambda)}\frac{a_+(\lambda_1,\lambda)}{b_-(\lambda_1,\lambda)} B(\lambda_3) B(\lambda_2)B(\lambda_1) A(\lambda) \nonumber \\
&- \frac{c_+(\lambda_3,\lambda)}{b_-(\lambda_3,\lambda)}\frac{a_+(\lambda_2,\lambda_3)}{b_-(\lambda_2,\lambda_3)}\frac{a_+(\lambda_1,\lambda_3)}{b_-(\lambda_1,\lambda_3)}B(\lambda)B(\lambda_2)B(\lambda_1)A(\lambda_3) \nonumber\\
&-\frac{c_+(\lambda_2,\lambda) }{b_-(\lambda_2,\lambda)}\frac{a_-(\lambda_3,\lambda_2)}{b_-(\lambda_3,\lambda_2)} \frac{a_+(\lambda_1,\lambda_2)}{b_-(\lambda_1,\lambda_2)} B(\lambda)B(\lambda_3)B(\lambda_1)A(\lambda_2)  \label{A3} \\
&+\Big[-\frac{c_+(\lambda_1,\lambda)}{b_-(\lambda_1,\lambda)}\frac{a_+(\lambda_3,\lambda)}{b_-(\lambda_3,\lambda)}\frac{a_+(\lambda_2,\lambda)}{b_-(\lambda_2,\lambda)} \overbrace{B(\lambda_3)\underbrace{B(\lambda_2)B(\lambda)}_{eq. (\ref{alg1})}}^{eq. (\ref{alg1})} \nonumber \\
&+\frac{c_+(\lambda_3,\lambda)}{b_-(\lambda_3,\lambda)}\frac{c_+(\lambda_1,\lambda_3)}{b_-(\lambda_1,\lambda_3)}\frac{a_+(\lambda_2,\lambda_3)}{b_-(\lambda_2,\lambda_3)} B(\lambda)\underbrace{B(\lambda_2)B(\lambda_3)}_{eq. (\ref{alg1})} \nonumber \\
&+\frac{c_+(\lambda_2,\lambda)}{b_-(\lambda_2,\lambda)}\underbrace{\frac{c_+(\lambda_1,\lambda_2)}{b_-(\lambda_1,\lambda_2)}}_{eq. (\ref{unit1})}\frac{a_-(\lambda_3,\lambda_2)}{b_-(\lambda_3,\lambda_2)} B(\lambda)B(\lambda_3)B(\lambda_2) \Big]A(\lambda_1). \nonumber
\end{alignat}
Again one should use the commutation rule (\ref{alg1}) and unitarity relation (\ref{unit1}) as indicated. So the last term of (\ref{A3}) can be rewritten as follows,
\bear
-\Big[\frac{c_+(\lambda_1,\lambda)}{b_-(\lambda_1,\lambda)}\frac{a_-(\lambda_3,\lambda)}{b_-(\lambda_3,\lambda)}\frac{a_-(\lambda_2,\lambda)}{b_-(\lambda_2,\lambda)} -\frac{c_+(\lambda_3,\lambda)}{b_-(\lambda_3,\lambda)}\frac{c_+(\lambda_1,\lambda_3)}{b_-(\lambda_1,\lambda_3)}\frac{a_-(\lambda_2,\lambda_3)}{b_-(\lambda_2,\lambda_3)}  \nonumber \\
+\frac{c_+(\lambda_2,\lambda)}{b_-(\lambda_2,\lambda)}\frac{c_-(\lambda_2,\lambda_1)}{b_-(\lambda_2,\lambda_1)}\frac{a_-(\lambda_3,\lambda_2)}{b_-(\lambda_3,\lambda_2)}\Big] B(\lambda)B(\lambda_3)B(\lambda_2) A(\lambda_1). \label{II1}
\ear
At this point, we have to simplify the term inside the square bracket ($I_1$). This can be done by joining the first and the last term on (\ref{II1}) and using  the equation (\ref{eq10}) a couple of times, such that
\begin{alignat}{1}
&I_1=\frac{[ b_-(\lambda_2,\lambda_1) a_-(\lambda_2,\lambda) c_+(\lambda_1,\lambda) ]a_-(\lambda_3,\lambda) b_-(\lambda_3,\lambda_2)} {b_-(\lambda_1,\lambda) b_-(\lambda_3,\lambda) b_-(\lambda_2,\lambda) b_-(\lambda_2,\lambda_1) b_-(\lambda_3,\lambda_2)} \nonumber\\
&+ \frac{\overbrace{ [a_-(\lambda_3,\lambda_2)b_-(\lambda_3,\lambda) c_+(\lambda_2,\lambda)]}^{eq. (\ref{eq10})} c_-(\lambda_2,\lambda_1) b_-(\lambda_1,\lambda) } {b_-(\lambda_1,\lambda) b_-(\lambda_3,\lambda) b_-(\lambda_2,\lambda) b_-(\lambda_2,\lambda_1) b_-(\lambda_3,\lambda_2)} \\
&-\frac{c_+(\lambda_3,\lambda)}{b_-(\lambda_3,\lambda)}\frac{c_+(\lambda_1,\lambda_3)}{b_-(\lambda_1,\lambda_3)}\frac{a_-(\lambda_2,\lambda_3)}{b_-(\lambda_2,\lambda_3)}, \nonumber
\end{alignat}
and
\begin{alignat}{1}
&I_1=\frac{ a_-(\lambda_3,\lambda)}{ b_-(\lambda_3,\lambda)}\frac{\overbrace{[ b_-(\lambda_2,\lambda_1) a_-(\lambda_2,\lambda) c_+(\lambda_1,\lambda) + c_-(\lambda_2,\lambda_1) c_+(\lambda_2,\lambda) b_-(\lambda_1,\lambda)]}^{eq. (\ref{eq10})} }{b_-(\lambda_2,\lambda_1)  b_-(\lambda_2,\lambda) b_-(\lambda_1,\lambda) } \nonumber\\
&+ \underbrace{\frac{c_-(\lambda_3,\lambda_2)}{b_-(\lambda_3,\lambda_2)}}_{eq. (\ref{unit1})}\frac{c_+(\lambda_3,\lambda)} {b_-(\lambda_3,\lambda)}\frac{c_-(\lambda_2,\lambda_1)}{  b_-(\lambda_2,\lambda_1)} 
-\frac{c_+(\lambda_3,\lambda)}{b_-(\lambda_3,\lambda)}\frac{c_+(\lambda_1,\lambda_3)}{b_-(\lambda_1,\lambda_3)}\frac{a_-(\lambda_2,\lambda_3)}{b_-(\lambda_2,\lambda_3)}. 
\end{alignat}
Then we again have to use the equation (\ref{eq10}) twice,
\begin{alignat}{1}
&I_1=\frac{c_+(\lambda_1,\lambda)}{b_-(\lambda_1,\lambda) }\frac{ a_-(\lambda_3,\lambda)}{ b_-(\lambda_3,\lambda)}\frac{ a_-(\lambda_2,\lambda_1)}{b_-(\lambda_2,\lambda_1)} \\
&-\frac{c_+(\lambda_3,\lambda)} {b_-(\lambda_3,\lambda)}  \frac{\overbrace{[c_-(\lambda_2,\lambda_1)c_+(\lambda_2,\lambda_3) b_-(\lambda_1,\lambda_3) +  b_-(\lambda_2,\lambda_1) a_-(\lambda_2,\lambda_3)c_+(\lambda_1,\lambda_3)]}^{eq. (\ref{eq10})} }{  b_-(\lambda_2,\lambda_1)b_-(\lambda_2,\lambda_3)b_-(\lambda_1,\lambda_3)}, \nonumber \\
&=\frac{ a_-(\lambda_2,\lambda_1)}{b_-(\lambda_2,\lambda_1)} \Big[ \frac{c_+(\lambda_1,\lambda)}{b_-(\lambda_1,\lambda) }\frac{ a_-(\lambda_3,\lambda)}{ b_-(\lambda_3,\lambda)} - \underbrace{\frac{  c_+(\lambda_1,\lambda_3) }{b_-(\lambda_1,\lambda_3)}}_{eq. (\ref{unit1})}\frac{c_+(\lambda_3,\lambda)} {b_-(\lambda_3,\lambda)} \Big],
\end{alignat}
and
\begin{alignat}{1}
I_1&=\frac{ a_-(\lambda_2,\lambda_1)}{b_-(\lambda_2,\lambda_1)}  \frac{ \overbrace{[b_-(\lambda_3,\lambda_1) a_-(\lambda_3,\lambda) c_+(\lambda_1,\lambda)+ c_-(\lambda_3,\lambda_1) c_+(\lambda_3,\lambda)b_-(\lambda_1,\lambda) ]}^{eq. (\ref{eq10})}}{ b_-(\lambda_3,\lambda_1) b_-(\lambda_3,\lambda)b_-(\lambda_1,\lambda) }, \\
&=\frac{a_-(\lambda_2,\lambda_1)}{b_-(\lambda_2,\lambda_1)}  \frac{a_-(\lambda_3,\lambda_1)}{ b_-(\lambda_3,\lambda_1)} \frac{c_+(\lambda_1,\lambda)}{b_-(\lambda_1,\lambda) }.
 \label{FI1}
\end{alignat}
Finally, we can substitute the final result for the term $I_1$ (\ref{FI1}) in Eq.(\ref{A3}), 
\begin{alignat}{1}
&A(\lambda)B(\lambda_3)B(\lambda_2)B(\lambda_1)=\frac{a_+(\lambda_3,\lambda)}{b_-(\lambda_3,\lambda)}\frac{a_+(\lambda_2,\lambda)}{b_-(\lambda_2,\lambda)}\frac{a_+(\lambda_1,\lambda)}{b_-(\lambda_1,\lambda)} B(\lambda_3) B(\lambda_2)B(\lambda_1) A(\lambda) \nonumber \\
&- \frac{c_+(\lambda_3,\lambda)}{b_-(\lambda_3,\lambda)}\frac{a_+(\lambda_2,\lambda_3)}{b_-(\lambda_2,\lambda_3)}\frac{a_+(\lambda_1,\lambda_3)}{b_-(\lambda_1,\lambda_3)}B(\lambda)B(\lambda_2)B(\lambda_1)A(\lambda_3) \nonumber\\
&-\frac{c_+(\lambda_2,\lambda) }{b_-(\lambda_2,\lambda)}\frac{a_-(\lambda_3,\lambda_2)}{b_-(\lambda_3,\lambda_2)} \frac{a_+(\lambda_1,\lambda_2)}{b_-(\lambda_1,\lambda_2)} B(\lambda)B(\lambda_3)B(\lambda_1)A(\lambda_2)  \label{A3B} \\
&-\frac{c_+(\lambda_1,\lambda)}{b_-(\lambda_1,\lambda) }\frac{a_-(\lambda_3,\lambda_1)}{ b_-(\lambda_3,\lambda_1)} \frac{a_-(\lambda_2,\lambda_1)}{b_-(\lambda_2,\lambda_1)}   B(\lambda)B(\lambda_3)B(\lambda_2) A(\lambda_1). \nonumber
\end{alignat}
This expression coincides with the formula (\ref{AoverB}) for the case $M=3$. Note that we have only used the commutation rules (\ref{alg1},\ref{alg2}), the unitarity relation (\ref{unit1}) and the Yang-Baxter relation (\ref{eq10}). It is remarkable that the same applies to any $M$-values, where we again would need only the mentioned relations.

Similarly, the expression (\ref{DoverB}) can be obtained using the commutation rules (\ref{alg1},\ref{alg3}), the unitarity relation (\ref{unit1}) and the Yang-Baxter relation (\ref{eq2}).

Alternatively, one could have obtained the formula (\ref{AoverB}) in a shorter way. Note that the first term in (\ref{AoverB}) is obtained straightforwardly using the first term of the algebraic relation (\ref{alg2}) $M$-times. On the other hand, the second term of (\ref{AoverB}) can be very complicated, as we have seen above. The coefficient of the term not containing the first $B$-operator, $B(\lambda_M)$, is very simple though. This coefficient is obtained using the second term of (\ref{alg2}) in order to exchange the $A(\lambda)$ and $B(\lambda_M)$ and in all other steps one has to use only the first term of (\ref{alg2}) \cite{QISM-rev}, which results in the following expression
\begin{alignat}{1}
&A(\lambda,\{\nu_k\})\prod_{i=1}^M B(\lambda_i,\{\nu_k\})=\prod_{i=1}^M\frac{a_+(\lambda_i,\lambda)}{b_-(\lambda_i,\lambda)} \prod_{i=1}^M B(\lambda_i,\{\nu_k\}) A(\lambda,\{\nu_k\}) \nonumber\\
&- \frac{c_+(\lambda_M,\lambda)}{b_-(\lambda_M,\lambda)} \prod_{\stackrel{i=1}{i\neq M}}^M \frac{a_+(\lambda_i,\lambda_j)}{b_-(\lambda_i,\lambda_j)} B(\lambda,\{\nu_k\}) \prod_{\stackrel{i=1}{i\neq M}}^M B(\lambda_i,\{\nu_k\}) A(\lambda_M,\{\nu_k\}) \label{partialAoB}\\
&+ \mbox{sum of other terms not containing $B(\lambda_j)$, $j=1,\cdots,M-1$}. \nonumber
\end{alignat}

However, one could have done similar analysis for any $B(\lambda_j)$. In doing so, one has to use the relation (\ref{alg1}) in order to move $B(\lambda_j)$ to the leftmost position, 
\bear
B(\lambda_M) \cdots \underbrace{B(\lambda_{j+1}) B(\lambda_{j})}_{eq. (\ref{alg1})} \cdots B(\lambda_1)= B(\lambda_j) \prod_{i=1 \atop i\neq j}^{M} B(\lambda_i)\theta_{>}(\lambda_i,\lambda_j)
\ear
and then repeat the above described procedure. This would result in the missing terms in (\ref{partialAoB}) and finally we should obtain the expression (\ref{AoverB}).

\end{document}